\documentclass[a4wide,12pt,showpacs,tightenlines,preprint,titlepage,nofootinbib,aps]{revtex4}
\bibstyle{unsrt}

\newcommand{\half}{\frac{1}{2}}
\newcommand{\quarter}{\frac{1}{4}}
\newcommand{\bea}{\begin{eqnarray}}
\newcommand{\eea}{\end{eqnarray}}
\newcommand{\vf}{\varphi}
\begin{document}

\title{The disappearing $Q$ operator}
\author{H.~F.~Jones and R.~J.~Rivers}
\affiliation{Physics Department, Imperial College, London, SW7 2AZ, UK\\}
\date{\today\\}
\begin{abstract}$\phantom{x}$\\
In the Schr\"odinger formulation of non-Hermitian quantum theories a positive-definite
metric operator $\eta\equiv e^{-Q}$ must be introduced in order to ensure their
probabilistic interpretation. This operator also gives an equivalent Hermitian theory,
by means of a similarity transformation. If, however, quantum mechanics is formulated in terms of
functional integrals, we show that the $Q$ operator makes only a subliminal appearance and is not needed
for the calculation of expectation values. Instead, the relation to the Hermitian theory is
encoded via the external source $j(t)$. These points are illustrated and amplified for two
non-Hermitian quantum theories: the Swanson model, a non-Hermitian transform of the simple harmonic oscillator,
and the wrong-sign quartic oscillator, which has been shown to be equivalent to a conventional asymmetric
quartic oscillator.
\end{abstract}
\maketitle

\section{Introduction}
The recent interest in non-Hermitian quantum mechanics stems from the seminal paper
by Bender and Boettcher\cite{BB}, who showed numerically that Hamiltonians of the form
\bea
H=\half p^2-g (ix)^N
\eea
possessed a real, positive spectrum for $N\ge 2$. This reality was attributed to an
unbroken $PT$ symmetry of $H$ and was subsequently proved analytically by Dorey et al.\cite{Dorey},
exploiting the connection between ordinary differential equations and
integrable models.

The more general framework of pseudo-Hermiticity, encompassing $PT$ symmetry, was subsequently formulated
by Mostafazadeh\cite{AM}, whereby
\bea\label{pseud}
H^\dag=\eta H \eta^{-1},
\eea
in which $\eta$ is a positive-definite, Hermitian operator. Essentially the same idea had appeared
previously under the name of ``quasi-Hermiticity" by Scholtz et al.\cite{Scholtz}, and indeed goes
back to much earlier work by Pauli\cite{Pauli}.

Reality of the eigenvalues is not sufficient for a viable non-Hermitian quantum theory: for a
probabilistic physical interpretation we must also have a positive-definite Hilbert-space metric.
Such a metric may be constructed, but it is not given a priori, and depends on the particular
Hamiltonian $H$. In the context of $PT$ symmetry a new operator, the $C$ operator, was introduced
in Ref.~\cite{Cop}, with the help of which a positive-definite, $CPT$ inner product could be constructed.
In a subsequent paper\cite{Qop} a systematic way of constructing $C$ in perturbation theory was developed, which
was greatly facilitated by the introduction of the $Q$ operator, defined by $PC=e^{-Q}$. The connection
with the operator $\eta$ of Eq.~(\ref{pseud}) is
\bea
\eta=e^{-Q}.
\eea
This appears in the overlap $\langle\langle\ , \ \rangle\rangle$, defined by
\bea\label{overlap}
\langle\langle\psi,\vf\rangle\rangle\equiv \langle\psi, e^{-Q}\vf\rangle,
\eea
and in the expression for the equivalent Hermitian Hamiltonian\cite{AM}:
\bea
h=e^{-\half Q}He^{\half Q},
\eea
showing that the two operators are related by a similarity transformation (rather than a
unitary transformation).

In most cases one must resort to perturbation theory or some other approximation in order to
construct $Q$ and $h$. However, in the two theories we study below they are known exactly.
These are the Swanson\cite{Swanson} Hamiltonian,
\bea\label{Swanson}
H=\omega a^\dag a +\alpha a^2 + \beta {a^\dag}^2,
\eea
where $a$ and $a^\dag$ are simple harmonic oscillator annihilation and
creation operators for unit frequency and $\omega$, $\alpha$ and $\beta$ are real,
and the ``wrong-sign" quartic oscillator,
\bea\label{WS}
H=\half p_z^2 +\half m^2z^2 -gz^4,
\eea
where we have written $z$, $p_z$ in place of $x$, $p$ in recognition of the fact that for the
energy eigenvalue equation to be well defined, it must be formulated on a contour in the lower-
half complex $z$ plane.

The $Q$ operator (or operators, because it is not unique) can be found exactly\cite{Geyer, HFJ} for
Eq.~(\ref{Swanson}), and the equivalent Hermitian Hamiltonian is simply a harmonic oscillator.
In the case of Eq.~(\ref{WS}) it can again be found exactly for a particular choice of contour\cite{JM},
leading to an equivalent Hermitian Hamiltonian which is a standard quartic oscillator, together with
a linear term, as discovered earlier by Andrianov\cite{And} and Buslaev and Grecchi\cite{BG}.

Apart from the standard formulation of quantum mechanics in terms of operators $\hat{x}$ and $\hat{p}$
acting on Schr\"odinger wave functions we may also approach it from the point of view of functional
integrals, which indeed is the natural language if a generalization to higher dimensions is to be
attempted. A brief discussion of the Swanson model in this context was given in Ref.~\cite{Bologna} and
a rederivation of the results of Ref.~\cite{JM} using functional methods was given in \cite{CMBfi},
subsequently corrected in \cite{HFJfi}. These papers were primarily concerned with the derivation of the
equivalent Hermitian Hamiltonian $h$ and did not address the question of overlaps or expectation values.
Indeed, the quantity considered was the partition function, i.e. the vacuum-generating functional for
$j\equiv 0$, rather than the true functional $Z[j]$, from which Green functions can be obtained by functional derivation.

In the following two sections we will explore the full vacuum generating functional, finding that in this
formalism the operator $Q$ appears only fleetingly and is not involved in the calculation of overlaps or
expectation values. It is rather the dependence of $Z[j]$ on $j$ that encodes the similarity transformation
induced by $Q$. In passing, we also note the role of $Q$ at the classical level as being a generator of canonical
transformations. In the final section we discuss the implications of our results.

\section{$Z[j]$ for the Swanson Model}

In terms of $x$ and $p$ the Hamiltonian of Eq.~(\ref{Swanson}) reads
\bea\label{Swansonp}
H=a x^2+b p^2+c \{x,p\},
\eea
where $a=\half(\omega+\alpha+\beta)$, $b=\half(\omega-\alpha-\beta)$ and $c=\half i(\alpha-\beta)$.
It is  non-Hermitian for $\alpha\ne\beta$.

The Euclidean phase-space functional integral for $Z[j]$ corresponding to Eq.~(\ref{Swansonp}) is
\bea\label{EPF1}
Z[j]=\int [D\vf][D\pi] \exp \Big\{ -\int dt[i\dot{\vf}\pi+a\vf^2+b\pi^2+2c\vf\pi-j\vf]\Big\},
\eea
where, to correspond to the notation in field theory, we have written $\vf$ for $x$ and
$\pi$ for $p$.

The different solutions for $Q$ and $h$ are quite simply obtained by completing the square in the
integrand in different ways\footnote{Below we have illustrated the two extreme cases where $Q=Q(\vf)$
and $Q=Q(\pi)$. The more general case $Q=Q(\vf, \pi)$, in which both $\vf$ and $\pi$ are shifted,
does not add anything essential to the case $Q=Q(\pi)$.}
\subsection{$Q=Q(\vf)$}\label{Qphi}
This corresponds to completing the square by shifting $\pi$, according to
\bea
a\vf^2+b\pi^2+2c\vf\pi=b\big(\pi+\frac{c}{b}\vf\big)^2+\big(a-\frac{c^2}{b}\big)\vf^2
\equiv b\tilde{\pi}^2+\tilde{a}\vf^2,
\eea
giving
\bea\label{ZSwanson}
Z[j]&=&\int [D\vf][D\tilde{\pi}] \exp \Big\{ -\int dt[i\dot{\vf}(\tilde{\pi}-
\frac{c}{b}\vf)+\tilde{a}\vf^2+b\tilde{\pi}^2-j\vf]\Big\}\cr&&\cr
&=&\int [D\vf][D\tilde{\pi}] \exp \Big\{ -\int dt[i\dot{\vf}\tilde{\pi}
+\tilde{a}\vf^2+b\tilde{\pi}^2-j\vf]\Big\},
\eea
The $Q(x)$ operator has previously been determined in the operator formalism\cite{HFJ} as
\bea
Q(x)=\frac{\alpha-\beta}{\omega-\alpha-\beta}x^2=-i\frac{c}{2b}x^2
\eea
Its time derivative, written in terms of $\vf$, appears in the first line of Eq.~(\ref{ZSwanson}),
but then disappears in the following line, precisely because it is a total derivative. The momentum
variable can now be integrated out\footnote{Here, and in subsequent integrations, we ignore overall
multiplicative factors.}, to leave
\bea\label{ZSHO}
Z[j]=\int [D\vf] \exp \Big\{ -\int dt\ \Big[\frac{\dot{\vf}^2}{4b}
+\tilde{a}\vf^2-j\vf\Big]\Big\},
\eea
the standard form for the vacuum generating functional of the simple harmonic oscillator
with frequency $\Omega^2\equiv 4 \tilde{a}b$. Note that by completing the square in this particular
way, we have effectively changed $\pi$ to $\tilde{\pi}\equiv \pi+(c/b)\vf$ but left $\vf$ unchanged.
In the operator formalism this would be the result of the transformations
\bea
p&\to& P\equiv e^{\half Q(x)}p\ e^{-\half Q(x)}=p+\half i Q^{\, \prime}(x),\nonumber\\
x&\to& X\equiv e^{\half Q(x)}x\ e^{-\half Q(x)}=x.
\eea
In terms of classical mechanics this is a canonical transformation, generated by
\bea
F_2(x,P)=xP-\half i Q(x)
\eea
according to the standard equations
\bea
X=\frac{\partial F_2}{\partial P}, \quad p=\frac{\partial F_2}{\partial x}.
\eea
It transforms the original non-Hermitian Hamiltonian $H$ of Eq.~(\ref{Swansonp}) into the equivalent Hermitian Hamiltonian
\bea
h= b P^2+\tilde{a} X^2,
\eea
while the corresponding
Lagrangians, $\ell$ and $L$, differ precisely by the time derivative of $Q$.

In this way of constructing $h$ the field $\vf$ remains an observable in the original theory,
and Green functions of $\vf$ are obtained in the standard way by functional differentiation
with respect to $j(t)$. Since the single-variable functional integral for $Z[j]$ obtained from
Eq.~(\ref{EPF1}) is identical that of Eq.~(\ref{ZSHO}), the Green functions of the Swanson Hamiltonian
can be obtained in the functional formalism without reference to $Q$, in contrast with the corresponding
calculations within the standard Schr\"odinger framework, where matrix elements have to be calculated
with the metric $\eta=e^{-Q}$, as in Eq.~(\ref{overlap}).

\subsection{$Q=Q(\pi)$}

In this case, there will be a shift in $\vf$ between the original and the Hermitian Hamiltonian.
For that reason, and to conform with the following section, we change the notation
so that the field appearing in Eq.~(\ref{EPF1}) becomes $\psi$:
\bea\label{EPF2}
Z[0]=\int [D\psi][D\pi] \exp \Big\{ -\int dt[i\dot{\psi}\pi+a\psi^2+b\pi^2+2c\psi\pi]\Big\},
\eea
reserving $\vf$ for the shifted field appearing in the Hermitian version. Note that for the
moment we have set $j=0$ in Eq.~(\ref{EPF2}). This is because we will ultimately wish to calculate
Green functions of the observable $\vf$, and hence will insert $j$ in the latter form and then
transform back to determine the $j$ dependence in Eq.~(\ref{EPF2}).
Completing the square according to
\bea
a\psi^2+b\pi^2+2c\psi\pi&=&a\big(\psi+\frac{c}{a}\pi\big)^2+\big(b-\frac{c^2}{a}\big)\pi^2\cr&&\cr
&\equiv& a\vf^2+\tilde{b}\pi^2,
\eea
we obtain
\bea\label{Z2Swanson}
Z[0]&=&\int [D\vf][D\pi] \exp \Big\{ -\int dt[i(\dot{\vf}-\frac{c}{a}\dot{\pi})\pi
+a\vf^2+\tilde{b}\pi^2]\Big\}\cr&&\cr
&=&\int [D\vf][D\pi] \exp \Big\{ -\int dt[i\dot{\vf}\pi
+a\vf^2+\tilde{b}\pi^2]\Big\},
\eea
after discarding the derivative of $Q(\pi)\equiv -i\pi^2(c/2a)$ appearing in the
first line of Eq.~(\ref{Z2Swanson}). Performing the integration over $\pi$ we obtain
\bea\label{ZSHO2}
Z[0]=\int [D\vf] \exp \Big\{ -\int dt\ \Big[\frac{\dot{\vf}^2}{4\tilde{b}}
+a\vf^2\Big]\Big\},
\eea
which is a rescaled version of Eq.~(\ref{ZSHO}) with $j=0$. The corresponding operator
transformations would be
\bea\label{Qp}
p&\to& P\equiv e^{\half Q(p)}p\ e^{-\half Q(p)}=p\nonumber\\
x&\to& X\equiv e^{\half Q(p)}x\ e^{-\half Q(p)}=x-\half i Q^{\, \prime}(p),
\eea
while the corresponding transformation in classical mechanics would be generated by
\bea
F_2(x,P)=xP-\half i Q(P).
\eea
In order to obtain Green functions of $\vf$ we now restore $j$ to Eq.~(\ref{ZSHO2}):
\bea\label{ZSHO3}
Z[j]=\int [D\vf] \exp \Big\{ -\int dt\ \Big[\frac{\dot{\vf}^2}{4\tilde{b}}
+a\vf^2-j\vf\Big]\Big\},
\eea
or equivalently in Eq.~(\ref{Z2Swanson}):
\bea
Z[j]&=&\int [D\vf][D\pi] \exp \Big\{ -\int dt[i\dot{\vf}\pi
+a\vf^2+\tilde{b}\pi^2-j\vf]\Big\},
\eea
and transform back to the original field variable $\psi$ by substituting
$\vf=\psi+(c/a)\pi$.
The resulting expression in the exponent:
\bea\nonumber
[\ ]=i\dot{\psi}\pi +b\pi^2 +a\psi^2+2c\psi\pi-j\psi-j(c/a)\pi
\eea
can be written in the form
\bea
[\ ]&=&b\left[\pi+\frac{1}{2b}\left(i\dot{\psi}+2c\psi-\frac{c}{a}j\right)\right]^2
+\frac{1}{4b}\dot{\psi}^2
+\tilde{a}\psi^2-\frac{\Omega^2}{4ab}j\psi
+\frac{ic}{2ab}j\dot{\psi}
-\frac{c^2}{4a^2b}j^2,
\eea
which gives
\bea\label{Z3Swanson}
Z[j]=\int [D\psi] \exp \Big\{ -\int dt\ \Big[\half\dot{\psi}^2
+\half\Omega^2\psi ^2-\frac{\Omega^2}{2a}\frac{j}{\sqrt{2b}}\psi
+\frac{ic}{a}\frac{j}{\sqrt{2b}}\dot{\psi}
-\frac{c^2j^2}{4a^2b}\Big]\Big\},
\eea
after performing the $\pi$ integration and rescaling $\psi$ by $\psi\to\psi\sqrt{2b}$.

Let us now verify that we obtain the correct expressions for some low-order Green functions
in $\vf$ by functional differentiation of Eq.~(\ref{Z3Swanson}).
A single differentiation $\delta/\delta j(t)$ gives
\bea
\left.\frac{1}{Z}\frac{\delta Z}{\delta j}\right|_{j=0}
=\frac{1}{\sqrt{2b}}\left\langle\frac{\Omega^2}{2a}\psi-\frac{ic}{a}\dot{\psi}\right\rangle,
\eea
which is clearly zero, the correct result for $\langle \vf \rangle$.

For a somewhat less trivial check we perform a double differentiation $\delta^2/\delta j_1\delta j_2$,
using an obvious subscript notation for $j$ at the two times $t_1$, $t_2$. Thus
\bea
\left.\frac{1}{Z}\frac{\delta^2 Z}{\delta j_1 \delta j_2}\right|_{j=0}
=\frac{1}{2b}\left\langle\left(\frac{\Omega^2}{2a}\psi_1-\frac{ic}{a}\dot{\psi_1}\right)
\left(\frac{\Omega^2}{2a}\psi_2-\frac{ic}{a}\dot{\psi_2}\right)\right\rangle +\frac{c^2}{2a^2b}\delta(t_1-t_2),
\eea
This is simply evaluated by noting that for the harmonic oscillator Lagrangian appearing in
Eq.~(\ref{Z3Swanson}) the expectation values are
\bea
\langle \psi_1 \psi_2 \rangle&=& \frac{1}{2\Omega}e^{-\Omega|t_1-t_2|}\nonumber\\
\langle \dot{\psi_1} \psi_2 \rangle=-\langle \psi_1\dot{\psi_2}\rangle
&=& \frac{1}{2}\varepsilon(t_2-t_1) e^{-\Omega|t_1-t_2|}\\
\langle \dot{\psi_1} \dot{\psi_2} \rangle
&=& -\frac{\Omega}{2}e^{-\Omega|t_1-t_2|}+\delta(t_1-t_2) \nonumber
\eea
The result is
\bea
\left.\frac{1}{Z}\frac{\delta^2 Z}{\delta j_1 \delta j_2}\right|_{j=0}
=\frac{\Omega}{4a}e^{-\Omega|t_1-t_2|},
\eea
after using the relation $c^2+\Omega^2/4=ab$. On the other hand, from Eq.~(\ref{ZSHO3}) we expect
\bea
\langle \vf_1 \vf_2 \rangle=\frac{\tilde{b}}{\Omega}e^{-\Omega|t_1-t_2|},
\eea
which is indeed the same because $\Omega^2=4a\tilde{b}$.

The lesson we draw from these simple calculations is that the Green functions of the observable
field $\vf$ appearing in the Lagrangian form of the equivalent Hermitian Hamiltonian $h$ can be calculated
using the functional integral arising from the non-Hermitian Hamiltonian $H$ without any reference to
the $Q$ operator, which is needed to calculate expectation values in the Schr\"odinger formulation of $H$.
Instead, the information about the transformation induced by $Q$ is encoded in the functional dependence
of $Z[j]$ on $j$.

\section{$Z[j]$ for $V(z)=-gz^4$}

As already mentioned, the Hamiltonian of Eq.~(\ref{WS}) gives rise to a well-defined eigenvalue
problem only if it is defined on a suitable contour in the lower-half complex $z$ plane. The essential
requirement on this contour is that it must lie asymptotically within the Stokes wedges\cite{BB}, which
in this case extend from the real axis down to an angle of $\pi/3$. Along the centre of the wedges
the wave-function behaves purely exponentially, changing to purely oscillatory at the edges. The particular
contour chosen in Ref.~\cite{JM},
\bea\label{eq:path}
z=-2i\surd(1+i\,x),
\eea
happens to go to infinity at an angle of $-\pi/4$, i.e. not down the centre of the wedge, but nonetheless
it has some very special properties that enable an exact evaluation of the $Q$ operator and the equivalent
Hermitian Hamiltonian. The results are that
\bea  \label{eq:q1}
Q=-\frac{p^3}{3\alpha}+2p\,,
\eea
where $\alpha\equiv 16 g$, and
\bea\label{WSh}
h=\frac{(p^2-4m^2)^2}{16\alpha}-\quarter p +\alpha x^2
\eea
Notice that this has the unusual feature that $p$ appears to the fourth power. However, by a Fourier transform
$h$ reduces to a standard, asymmetric quartic anharmonic oscillator. Note also that because $Q=Q(p)$
it induces the transformations of Eq.~(\ref{Qp}), whereby
\bea
P=p=-i\frac{d}{dx}=-\frac{2}{z}\frac{d}{dz}.
\eea
The result for $h$ has been reproduced in the functional integral formalism\cite{CMBfi, HFJfi}, but
again the quantity considered was $Z[0]$, not the full vacuum generating functional $Z[j]$. In the
present section we review the functional integral derivation, and include an external source in the
Hermitian formulation, tracing it back to the non-Hermitian functional integral.
\subsection{$Z[0]$}
We start with the single functional integral
\bea\label{Zorig}
Z=\int_C [D\psi] \exp \left\{-\int dt\left[\half\dot{\psi}^2+\half m^2\psi^2-g\psi^4\right]\right\},
\eea
defined, as discussed above, on the curve $C$ given by
\bea\label{cv}
\psi=-2i\surd(1+i\,\vf),
\eea
where $\vf$ is real.
Making the change of variable from $\psi$ to $\vf$, and including the additional effective potential
induced by this non-linear change\cite{JM,Lee}, we obtain
\bea\label{Zvf2}
Z[0]&=&\int \frac{[D\vf]}{{\rm Det}\surd(1+i\vf)} \exp \left\{-\int dt\left[\half\frac{\dot{\vf}^2}{1+i\vf}
-\frac{1}{32}\frac{1}{1+i\vf}-2m^2(1+i\vf)-\alpha(1+i\vf)^2\right]\right\}.\nonumber\\&&
\eea
Now we represent the functional determinant by a functional integral over $\pi$:
\bea
\frac{1}{{\rm Det}\surd(1+i\,\vf)}=\int [D\pi] \exp \left\{ -\int dt\,\half
(1+i\vf)\left(\pi-\frac{i\dot{\vf}+\frac{1}{4}}{1+i\vf}\right)^2\right\},
\eea
thus producing the
phase-space functional integral
\bea\label{psfi2}
Z[0]&=&\int [D\vf] [D\pi]\exp\left\{-\int dt \left[\half(1+i\vf)\pi^2-i\dot{\vf}\pi-\frac{\pi}{4}
-2m^2(1+i\vf)-\alpha(1+i\vf)^2\right]\right\}.\nonumber\\&&
\eea
Here the expression in square brackets in the exponent can be written as
\bea\label{sqbrack}
[\hspace{5mm}]=\alpha\left[(\vf-i)+\frac{i}{2\alpha}\left(\half(\pi^2-4m^4)+\dot{\pi}\right)\right]^2+\frac{\dot{\pi}^2}{4\alpha}
-\frac{\pi}{4}+\frac{(\pi^2-4m^2)^2}{16\alpha}.
\eea
So after rescaling $\vf\to\vf/\sqrt{2\alpha}$, $\pi\to \pi\sqrt{2\alpha}$ and performing
the $\vf$ integration we obtain
\bea
Z[0]=\int [D\pi] \exp\left\{-\int dt \left[ \frac{\dot{\pi}^2}{2}-\sqrt{\frac{\alpha}{8}}\pi
+\frac{\alpha}{4}\left(\pi^2-\frac{2m^2}{\alpha}\right)^2
\right]\right\},
\eea
in which (with $p\leftrightarrow \pi\sqrt{2\alpha}$) we can recognize the Lagrangian corresponding
to the Hermitian Hamiltonian $h$ of Eq.~(\ref{WSh}). Note again that in the transition to this last
equation $Q$ put in a brief appearance in the form of its derivative $\dot{Q}(\pi)$, which was immediately dropped.

\subsection{$Z[j]$}

Now let us couple $\pi$ to an external source $j$ by the addition of a term $-j\pi$ to $[\hspace{5mm}]$
and work backwards to see how $j$ appears in the the original functional integral.
In Eqs.~(\ref{sqbrack}) and (\ref{psfi2}) the term $\pi/4$ is replaced by
\bea
\frac{\pi}{4}\to \frac{\pi}{4}(1+4\tilde{j}),
\eea
where $\tilde{j}=j/\sqrt{2\alpha}$.
Then the $\pi$-dependent terms in the new version of Eq.~(\ref{psfi2}) can be written as
\bea
&&\hspace{-1.9cm}\half(1+i\vf)\pi^2-i\dot{\vf}\pi-\frac{\pi}{4}(1+4\tilde{j})\nonumber\\
&=&\half(1+i\vf)\left(\pi-\frac{i\dot{\vf}+\quarter(1+4\tilde{j})}{1+i\vf}\right)^2
+\half\frac{\dot{\vf}^2}{1+i\vf}-\frac{1}{32}\frac{1}{1+i\vf}-\frac{4i\dot{\vf}\tilde{j}+\tilde{j}+2\tilde{j}^2}{4(1+i\vf)}
\eea

So in the $[\hspace{5mm}]$ of Eq.~(\ref{Zvf2}) we will have the additional terms
\bea
-\frac{4i\dot{\vf}\tilde{j}+\tilde{j}+2\tilde{j}^2}{4(1+i\vf)}
\eea
Writing these in terms of $\psi$ using $4(1+i\vf)=-\psi^2$, so that
$i\dot{\vf}=-\half\psi\dot{\psi}$, we finally get
\bea\label{Zj}
Z[j]=\int_C [D\psi]\exp\left\{-\int dt \left[\half(\dot{\psi}^2+m^2\psi^2)-g\psi^4-
\tilde{j}\left(2\frac{\dot{\psi}}{\psi}-\frac{1}{\psi^2}\right)+\frac{2\tilde{j}^2}{\psi^2}\right] \right\}.
\eea

A functional differentiation with respect to $j$ gives
\bea\label{expv}
\left. \frac{1}{Z}\frac{\delta Z} {\delta j}\right|_{j=0}\equiv\langle \pi \rangle
=\frac{1}{\sqrt{2\alpha}}\left\langle 2\frac{\dot{\psi}}{\psi}-\frac{1}{\psi^2}\right\rangle.
\eea
Unfortunately, although we are assured that $H$ and $h$ are equivalent, with identical energy
spectra, we are unable to solve either exactly. Likewise we are not able to evaluate $\langle\pi\rangle$
exactly in either theory. However, in the next subsection we will check Eq.~(\ref{expv}) up to first
order in perturbation theory.

\subsection{Perturbation theory for $\langle\pi\rangle$}
\subsubsection{Hermitian formulation}

First we calculate $\langle\pi\rangle$ using normal Schr\"odinger perturbation theory for the Hermitian
Hamiltonian $h$ of Eq.~(\ref{WSh}). As a change of notation, let $p=y\sqrt{2\alpha}$,
so that $x=p_y/\sqrt{2\alpha}$, and $h$ now reads
\bea
h=\half p_y^2+\quarter\alpha\left(y^2-\frac{2m^2}{\alpha}\right)^2-y\sqrt{\frac{\alpha}{8}}
\eea
In order to perform perturbation theory in the standard way we need to shift $y$ according to
$y\to y+\beta$ so as to eliminate the linear term. This yields the cubic equation
\bea
\alpha\beta^3-2m^2\beta-\sqrt{\frac{\alpha}{8}}=0.
\eea
Although in principle it is possible to solve this equation exactly, it is extremely cumbersome
to do so: instead we will make successive approximations in powers of $\alpha$. In fact, the lowest-order
approximation is that $\beta$ is large, of order $1/\sqrt{\alpha}$. We thus write $\beta=\lambda/\sqrt{\alpha}$,
to obtain the equation
\bea
\lambda^3-2m^2\lambda=\frac{\alpha}{2\sqrt{2}},
\eea
whose solutions are approximately $\lambda=\pm m\sqrt{2}, \ -\alpha/(4m^2\sqrt{2})$. The overall
minimum is at the first of these, $\lambda=m\sqrt{2}$, and the lowest-order mass term is $\half M^2 y^2$,
where $M=2m$.
Thus the zeroth-order result for $\langle \pi\rangle$ is given by the classical position of the minimum,
i.e.
\bea\label{ClassicalResult}
\langle \pi\rangle=\beta=\frac{2m}{\sqrt{2\alpha}}\ .
\eea
In order to calculate the first-order result we need to do two things: first calculate the position of
the classical minimum to higher precision, and then calculate quantum corrections.

Calculating the minimum to next order we find that
\bea\label{FirstOrderShift}
\beta=\frac{2m}{\sqrt{2\alpha}}\left(1+\frac{g}{m^3}\right),
\eea
where we recall that $\alpha=16g$. The shifted Hamiltonian now reads
\bea
h=\half p_y^2+\half M^2 y^2 +m\sqrt{2\alpha}\left(1+\frac{g}{m^3}\right)y^3+\quarter \alpha y^4,
\eea
which gives a quantum contribution to $\langle y\rangle$ from a tadpole diagram arising from the
cubic coupling.  Its contribution is
\bea
\Delta\langle y \rangle = -3m\sqrt{2\alpha} \int dt' \frac{e^{-M|t-t'|}}{2M} \cdot \frac{1}{2M}
= -\frac{3m\sqrt{2\alpha}}{2M^3}
\eea
Combining this with the classical shift of Eq.~(\ref{FirstOrderShift}) we finally obtain
\bea
\langle \pi\rangle =\frac{2m}{\sqrt{2\alpha}}\left(1-\frac{2g}{m^3}\right).
\eea
\subsubsection{Functional integral formulation}

Now we wish to rederive these results from the functional integral of Eq.~(\ref{Zj}) and the expression
for $\langle\pi\rangle$ of Eq.~(\ref{expv}). In the latter equation a considerable simplification can
be made by noting that $\dot{\psi}/\psi$ is a perfect time derivative, and hence its expectation value vanishes,
since there is no preferred origin of time. The remaining term, $1/\psi^2$, might seem
singular, but we recall that the contour $C$ stays away from the origin. In fact $C$ is not suitable
for performing perturbation theory, because it is right on the edge of the Stokes wedge for the harmonic
oscillator. But, and this is the great advantage of the functional integral approach compared with the
Schr\"odinger approach to the non-Hermitian problem, the contour may be analytically
continued\footnote{This is not possible in the Schr\"odinger approach, because the calculation of
expectation values also involves the complex conjugate of the wave-function.} to a contour
compatible with the asymptotic requirements of both the harmonic oscillator and negative quartic oscillator
potentials and still avoiding the origin.

The calculation of $\langle\pi\rangle$ to lowest order can be achieved by the use of functional integration
by parts. We start from the identity
\bea\label{parts}
0&=&\int [d\psi]\frac{\delta}{\delta\psi(t_1)}\left[\frac{1}{\psi(t_2)}e^{-\half\int dt dt'
\psi(t)G_0^{-1}(t-t')\psi(t')}\right]\\
&=&\int [d\psi]\left[-\frac{\delta(t_1-t_2)}{\psi(t_1)^2}-\frac{1}{\psi(t_2)}\int dt' G_0^{-1}(t_1-t')\psi(t')
\right]e^{-\half\int dt dt'
\psi(t)G_0^{-1}(t-t')\psi(t')},
\eea
where $G_0(t-t')$ is the operator $(-\partial_t^2+m^2)\delta(t-t')$.
Integrating this with respect to $\int dt_1 G_0(t_2-t_1)$, we obtain
\bea
0=\int [d\psi]\left[-\frac{G_0(0)}{\psi(t_2)^2}-1\right]e^{-\half\int dt dt'
\psi(t)G_0^{-1}(t-t')\psi(t')}.
\eea
Since $G_0(t-t')=e^{-m|t-t'|}/(2m)$, this proves that
\bea\label{complex}
\left\langle \frac{1}{\psi^2}\right\rangle=-2m\ .
\eea
and hence $\langle\pi\rangle=2m/\sqrt{2\alpha}$, in accordance with Eq.~(\ref{ClassicalResult}).
This last equation, Eq.~(\ref{complex}), might seem rather odd, but it must always be remembered that $\psi$ is
complex.
In order to obtain the first-order correction, we must keep the interaction term in the exponent.
The analogue of Eq.~(\ref{parts}) is now
\bea
0&=&\int [d\psi]\frac{\delta}{\delta\psi(t_1)}\left[\frac{1}{\psi(t_2)}e^{-\half\int dt dt'
\psi(t)G_0^{-1}(t-t')\psi(t')+g\int dt \psi^4}\right]\\
&=&\int [d\psi]\left[-\frac{\delta(t_1-t_2)}{\psi(t_1)^2}-\frac{1}{\psi(t_2)}\int dt' G_0^{-1}(t_1-t')\psi(t')
+4g\psi(t_1)^3\right]e^{\dots}
\eea
Now integrate $\int dt_1 G_0(t_2-t_1)$ to obtain
\bea
0=\int [d\psi]\left[-\frac{G_0(0)}{\psi(t_2)^2}-1+\frac{4g}{\psi(t_2)}\int dt_1 G_0(t_2-t_1)\psi(t_1)^3 \right]
e^{\dots}
\eea
i.e.
\bea
0=\int [d\psi]\left[-\frac{1}{\psi(t_2)^2}-2m+\frac{4g}{\psi(t_2)}\int dt_1 e^{-m|t_2-t_1|}\psi(t_1)^3 \right]
e^{\dots}
\eea
Given the fact that in the Hermitian sector we found that $2m \to 2m -4g/m^2$, we need to show that
\bea\label{correct}
I\equiv\int dt_1 e^{-m|t_1-t_2|}\left\langle\frac{1}{\psi(t_2)} \psi(t_1)^3 \right\rangle=\frac{1}{m^2}.
\eea
This is an unfamiliar kind of expectation value, but we can represent it in terms of more familiar
quantities by using the integral representation
\bea
\frac{i}{\psi(t_2)-i\varepsilon}=\int_0^\infty d\lambda e^{i\lambda\psi(t_2)}
\eea
In fact, precisely because our contour does not go through the origin, we may drop the $i\varepsilon$
in the calculation of $I$, which now reads
\bea
I=-i \int dt_1 e^{-m|t_1-t_2|}\int_0^\infty \sum_{n=0}^\infty \frac{(i\lambda)^{2n+1}}
{(2n+1)!}\left\langle \psi_2^{2n+1} \psi_1^3 \right\rangle.
\eea
Thinking graphically in terms of contractions, these can occur in two ways:
\begin{itemize}
\item[(a)] $ \langle\psi_1 \psi_1\rangle \langle\psi_1 \psi_2\rangle \langle\psi_2 \psi_2\rangle^n$\\
This occurs with coefficient $3(2n+1)!!$ and  value $e^{-m|t_1-t_2|}/(2m)^{n+2}$
\item[(b)]$\langle\psi_1 \psi_2\rangle^4 \langle\psi_1 \psi_2\rangle^{n-1}$ \\
This occurs with coefficient $2n(2n+1)!!$ and  value $e^{-3m|t_1-t_2|}/(2m)^{n+2}$
\end{itemize}
Thus
\bea
I_a=\frac{3}{4m^2}\int_0^\infty dt_1 e^{-2m|t_1-t_2|}\int_0^\infty d\lambda\ \lambda
e^{-\lambda^2/(4m)}=\frac{3}{2m^2}\ ,
\eea
while
\bea
I_b=-\frac{1}{8m^3}\int_0^\infty dt_1 e^{-4m|t_1-t_2|}\int_0^\infty d\lambda\ \lambda^3
e^{-\lambda^2/(4m)}=-\frac{1}{2m^2}\ .
\eea
Thus $I\equiv I_a+I_b=1/m^2$, in agreement with Eq.~(\ref{correct}).

An alternative way of proceeding, which of course gives the same result, is to treat $I$ in terms of a vacuum expectation value of
a time-ordered product of operators and insert a complete set of states. The states involved
are only the first and third excited states, and the required matrix elements of $1/\psi_2$
can be calculated from those of $\psi_2$ by writing $\langle (\psi_2)^{2r}\rangle =
\langle (\psi_2)^{2r+1} \times (1/\psi_2)\rangle$ for $r=0,1$
and inserting the requisite number of intermediate states.

\section{Discussion}

The main thrust of this paper has been to illustrate how the metric, in the guise of the $Q$
operator, makes only a fleeting appearance when the theory is formulated in terms of functional
integrals rather than the operator formalism. Green functions are calculated as functional integrals
in the normal way. The information about observables, namely the transforms of Hermitian operators in the equivalent
Hermitian formulation, is carried in the non-Hermitian version by the functional dependence of
$Z[j]$ on the external current coupled to those observables in the latter formulation. In the case of
the wrong-sign quartic oscillator, there is in fact a technical advantage in the functional formulation
compared with the normal Schr\"odinger formulation, namely that the original contour can be distorted so
as to make perturbative calculations of Green function possible.

However, it must be admitted that the number of non-Hermitian theories, regarded as fundamental theories,
that are interesting and tractable, is at the moment rather limited. One other theory that has been the
subject of extensive investigation is the $igx^3$ potential. In this case one can stay on the real line,
but the $Q$ operator, and hence the equivalent Hermitian Hamiltonian $h$, is only known in low-order perturbation
theory\cite{hequiv, AMh}. It has been shown that perturbative calculations of the energy
levels are considerably easier in the context of the non-Hermitian theory, regarded as a field theory in
1 dimension, than in the equivalent Hermitian theory, where the potential in $h$ contains a number of
momentum-dependent terms\cite{CMBen}. However, calculations of Green functions are likely to be difficult in both
versions\cite{AMhh}, and indeed by the methods of the present paper, because the transformations
induced by $Q$ are complicated, with $Q$ being a function of both $x$ and $p$.

An alternative way of viewing non-Hermitian theories is as effective theories, which happen to be useful
in deriving tractable results, the prime example being the non-unitary Dyson mapping used by
Snyman and Geyer\cite{Snyman} to convert the Hermitian Richardson Hamiltonian into an equivalent
non-Hermitian boson-fermion Hamiltonian.

Nevertheless, the $-gz^4$ potential is still of great interest as a fundamental theory because of its
possible generalization to a 4-dimensional $-g\vf^4$ theory. The reason is that the indications in
perturbation theory are that such a theory could be asymptotically free, as was already stressed in
the original paper of Bender and Boettcher\cite{BB}, and spelt out in somewhat more detail in
Ref.~\cite{Bologna}. The first attempts at a generalization to higher dimensions were made in Ref.~\cite{CMBfi},
but it remains a challenging problem.

\end{document}